%% file: draft.tex
\documentclass[preprint,prd,tightenlines,superscriptaddress,nofootinbib]{revtex4-1}
\input{definitions}

\def\belletwo {Belle~II}
\def\belle {Belle}

\newcommand{\tev}{\ensuremath{\mathrm{\,Te\kern -0.1em V}}\xspace}
\newcommand{\gev}{\ensuremath{\mathrm{\,Ge\kern -0.1em V}}\xspace}
\newcommand{\mev}{\ensuremath{\mathrm{\,Me\kern -0.1em V}}\xspace}
\newcommand{\kev}{\ensuremath{\mathrm{\,ke\kern -0.1em V}}\xspace}
\newcommand{\ev}{\ensuremath{\mathrm{\,e\kern -0.1em V}}\xspace}
\newcommand{\gevc}{\ensuremath{{\mathrm{\,Ge\kern -0.1em V\!/}c}}\xspace}
\newcommand{\mevc}{\ensuremath{{\mathrm{\,Me\kern -0.1em V\!/}c}}\xspace}
\newcommand{\gevcc}{\ensuremath{{\mathrm{\,Ge\kern -0.1em V\!/}c^2}}\xspace}
\newcommand{\mevcc}{\ensuremath{{\mathrm{\,Me\kern -0.1em V\!/}c^2}}\xspace}

\def\invfb   {\ensuremath{\mbox{\,fb}^{-1}}\xspace}

\def\epem       {\ensuremath{e^+e^-}\xspace}

\newcommand{\Mmin}{\ensuremath{M_{min}~}}

\newcommand{\mtau}{\ensuremath{m_{\tau}}}

\newcommand{\MCTauMassFit}{1777.72~\pm~0.17}

\newcommand{\MCEstimatorBias}{\ensuremath{0.72 \pm 0.12}}

\newcommand{\SystFitFunc}{0.08} 
\newcommand{\SystEstimator}{0.12}
\newcommand{\SystMomentumSF}{0.29}
\newcommand{\SystEstimatorMassDep}{0.02}
\newcommand{\SystWindow}{0.04} 
\newcommand{\SystBeamEnergy}{0.03}

\let\oldequation\equation
\let\oldendequation\endequation
\renewenvironment{equation}
  {\linenomathNonumbers\oldequation}
  {\oldendequation\endlinenomath}

\begin{document}


\includegraphics[width=3cm]{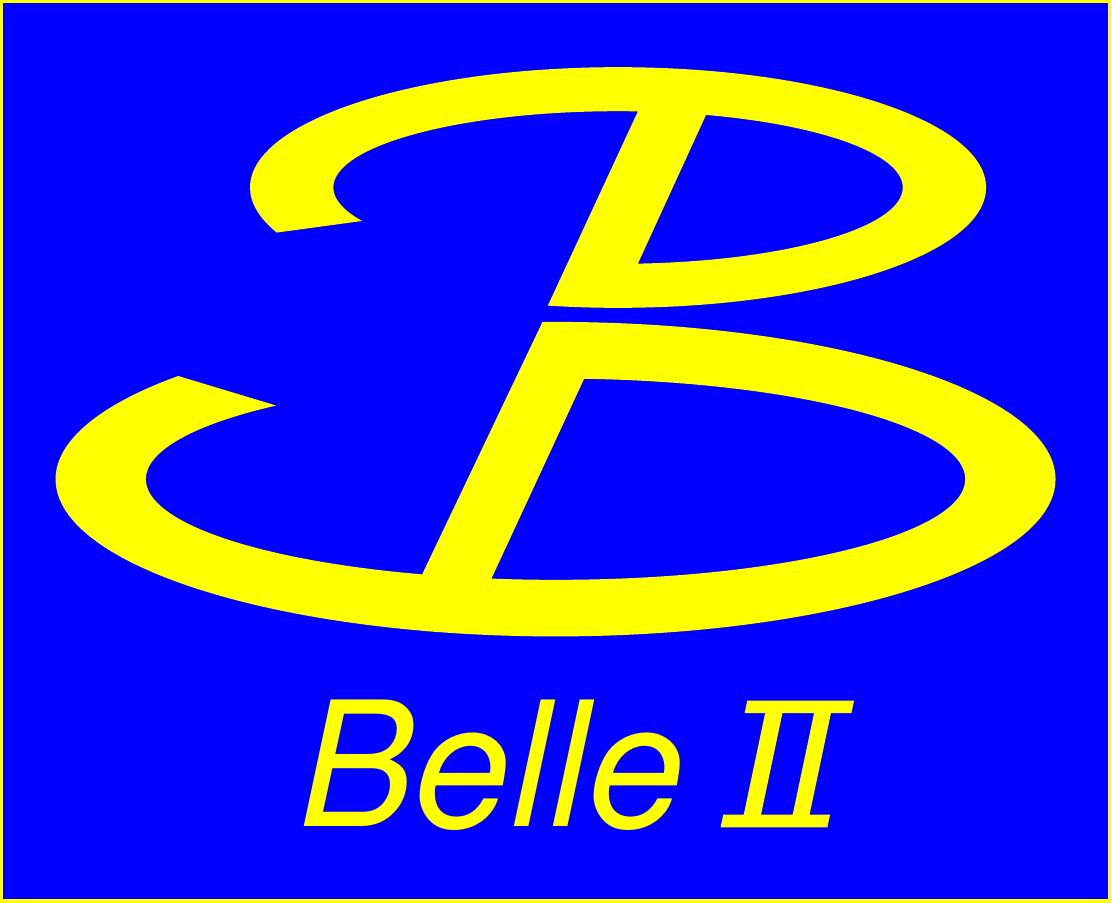}\vspace*{-3cm}

\begin{flushright}
    BELLE2-CONF-PH-2020-010\\
	\today
\end{flushright}
\vspace*{50pt}

\title{\LARGE $\tau$ lepton mass measurement at {\belletwo}}

\input{authors-conf2020.tex}


\begin{abstract} 
\input{abstract}
\end{abstract} 


\maketitle


\input{main}

\section*{Acknowledgements}
\input{acknowledgements}

\bibliography{references}

\end{document}

%% file: definitions.tex
\usepackage{graphicx} 
\usepackage{epstopdf} 
\usepackage{dcolumn} 
\usepackage{xcolor} 
\usepackage{amssymb} 
\usepackage{hyperref} 
\usepackage[utf8]{inputenc} 
\usepackage[english]{babel} 
\usepackage[displaymath, mathlines]{lineno} 
\usepackage{blindtext} 
\usepackage{amsmath}
\usepackage{gensymb}
\inputencoding{utf8}
\usepackage[T1]{fontenc}
\usepackage{ulem}

\graphicspath{{figures/}} 


%% file: authors-conf2020.tex
\newcommand{\instSinica}{Academia Sinica, Taipei 11529, Taiwan}
\newcommand{\instCPPM}{Aix Marseille Universit\'{e}, CNRS/IN2P3, CPPM, 13288 Marseille, France}
\newcommand{\instBeihang}{Beihang University, Beijing 100191, China}
\newcommand{\instBUAP}{Benemerita Universidad Autonoma de Puebla, Puebla 72570, Mexico}
\newcommand{\instBNL}{Brookhaven National Laboratory, Upton, New York 11973, U.S.A.}
\newcommand{\instBINP}{Budker Institute of Nuclear Physics SB RAS, Novosibirsk 630090, Russian Federation}
\newcommand{\instCMU}{Carnegie Mellon University, Pittsburgh, Pennsylvania 15213, U.S.A.}
\newcommand{\instCinvestavIPN}{Centro de Investigacion y de Estudios Avanzados del Instituto Politecnico Nacional, Mexico City 07360, Mexico}
\newcommand{\instPrague}{Faculty of Mathematics and Physics, Charles University, 121 16 Prague, Czech Republic}
\newcommand{\instChiangMai}{Chiang Mai University, Chiang Mai 50202, Thailand}
\newcommand{\instChiba}{Chiba University, Chiba 263-8522, Japan}
\newcommand{\instChonnam}{Chonnam National University, Gwangju 61186, South Korea}
\newcommand{\instConacyt}{Consejo Nacional de Ciencia y Tecnolog\'{\i}a, Mexico City 03940, Mexico}
\newcommand{\instDESY}{Deutsches Elektronen--Synchrotron, 22607 Hamburg, Germany}
\newcommand{\instDuke}{Duke University, Durham, North Carolina 27708, U.S.A.}
\newcommand{\instITAR}{Institute of Theoretical and Applied Research (ITAR), Duy Tan University, Hanoi 100000, Vietnam}
\newcommand{\instENEA}{ENEA Casaccia, I-00123 Roma, Italy}
\newcommand{\instEri}{Earthquake Research Institute, University of Tokyo, Tokyo 113-0032, Japan}
\newcommand{\instJuelich}{Forschungszentrum J\"{u}lich, 52425 J\"{u}lich, Germany}
\newcommand{\instFuJen}{Department of Physics, Fu Jen Catholic University, Taipei 24205, Taiwan}
\newcommand{\instFudan}{Key Laboratory of Nuclear Physics and Ion-beam Application (MOE) and Institute of Modern Physics, Fudan University, Shanghai 200443, China}
\newcommand{\instGoettingen}{II. Physikalisches Institut, Georg-August-Universit\"{a}t G\"{o}ttingen, 37073 G\"{o}ttingen, Germany}
\newcommand{\instGifu}{Gifu University, Gifu 501-1193, Japan}
\newcommand{\instSOKENDAI}{The Graduate University for Advanced Studies (SOKENDAI), Hayama 240-0193, Japan}
\newcommand{\instGyeongsang}{Gyeongsang National University, Jinju 52828, South Korea}
\newcommand{\instHanyang}{Department of Physics and Institute of Natural Sciences, Hanyang University, Seoul 04763, South Korea}
\newcommand{\instKEK}{High Energy Accelerator Research Organization (KEK), Tsukuba 305-0801, Japan}
\newcommand{\instJPARC}{J-PARC Branch, KEK Theory Center, High Energy Accelerator Research Organization (KEK), Tsukuba 305-0801, Japan}
\newcommand{\instHSE}{Higher School of Economics (HSE), Moscow 101000, Russian Federation}
\newcommand{\instIISER}{Indian Institute of Science Education and Research Mohali, SAS Nagar, 140306, India}
\newcommand{\instIITBhubaneswar}{Indian Institute of Technology Bhubaneswar, Satya Nagar 751007, India}
\newcommand{\instIITGuwahati}{Indian Institute of Technology Guwahati, Assam 781039, India}
\newcommand{\instIITHyderabad}{Indian Institute of Technology Hyderabad, Telangana 502285, India}
\newcommand{\instIITMadras}{Indian Institute of Technology Madras, Chennai 600036, India}
\newcommand{\instIndiana}{Indiana University, Bloomington, Indiana 47408, U.S.A.}
\newcommand{\instIHEPRussia}{Institute for High Energy Physics, Protvino 142281, Russian Federation}
\newcommand{\instHEPHYVienna}{Institute of High Energy Physics, Vienna 1050, Austria}
\newcommand{\instIHEPChina}{Institute of High Energy Physics, Chinese Academy of Sciences, Beijing 100049, China}
\newcommand{\instChennai}{Institute of Mathematical Sciences, Chennai 600113, India}
\newcommand{\instIPP}{Institute of Particle Physics (Canada), Victoria, British Columbia V8W 2Y2, Canada}
\newcommand{\instIOP}{Institute of Physics, Vietnam Academy of Science and Technology (VAST), Hanoi, Vietnam}
\newcommand{\instIFIC}{Instituto de Fisica Corpuscular, Paterna 46980, Spain}
\newcommand{\instFrascati}{INFN Laboratori Nazionali di Frascati, I-00044 Frascati, Italy}
\newcommand{\instNapoliINFN}{INFN Sezione di Napoli, I-80126 Napoli, Italy}
\newcommand{\instPadovaINFN}{INFN Sezione di Padova, I-35131 Padova, Italy}
\newcommand{\instPerugiaINFN}{INFN Sezione di Perugia, I-06123 Perugia, Italy}
\newcommand{\instPisaINFN}{INFN Sezione di Pisa, I-56127 Pisa, Italy}
\newcommand{\instRomaINFN}{INFN Sezione di Roma, I-00185 Roma, Italy}
\newcommand{\instRomaTreINFN}{INFN Sezione di Roma Tre, I-00146 Roma, Italy}
\newcommand{\instTorinoINFN}{INFN Sezione di Torino, I-10125 Torino, Italy}
\newcommand{\instTriesteINFN}{INFN Sezione di Trieste, I-34127 Trieste, Italy}
\newcommand{\instJAEA}{Advanced Science Research Center, Japan Atomic Energy Agency, Naka 319-1195, Japan}
\newcommand{\instMainz}{Johannes Gutenberg-Universit\"{a}t Mainz, Institut f\"{u}r Kernphysik, D-55099 Mainz, Germany}
\newcommand{\instGiessen}{Justus-Liebig-Universit\"{a}t Gie\ss{}en, 35392 Gie\ss{}en, Germany}
\newcommand{\instKarlsruhe}{Institut f\"{u}r Experimentelle Teilchenphysik, Karlsruher Institut f\"{u}r Technologie, 76131 Karlsruhe, Germany}
\newcommand{\instKennesaw}{Kennesaw State University, Kennesaw, Georgia 30144, U.S.A.}
\newcommand{\instKitasato}{Kitasato University, Sagamihara 252-0373, Japan}
\newcommand{\instKISTI}{Korea Institute of Science and Technology Information, Daejeon 34141, South Korea}
\newcommand{\instKorea}{Korea University, Seoul 02841, South Korea}
\newcommand{\instKSU}{Kyoto Sangyo University, Kyoto 603-8555, Japan}
\newcommand{\instKyotoU}{Kyoto University, Kyoto 606-8501, Japan}
\newcommand{\instKyungpook}{Kyungpook National University, Daegu 41566, South Korea}
\newcommand{\instLPI}{P.N. Lebedev Physical Institute of the Russian Academy of Sciences, Moscow 119991, Russian Federation}
\newcommand{\instLNNU}{Liaoning Normal University, Dalian 116029, China}
\newcommand{\instLMU}{Ludwig Maximilians University, 80539 Munich, Germany}
\newcommand{\instLuther}{Luther College, Decorah, Iowa 52101, U.S.A.}
\newcommand{\instMNITJaipur}{Malaviya National Institute of Technology Jaipur, Jaipur 302017, India}
\newcommand{\instMPP}{Max-Planck-Institut f\"{u}r Physik, 80805 M\"{u}nchen, Germany}
\newcommand{\instMPGHLL}{Semiconductor Laboratory of the Max Planck Society, 81739 M\"{u}nchen, Germany}
\newcommand{\instMcGill}{McGill University, Montr\'{e}al, Qu\'{e}bec, H3A 2T8, Canada}
\newcommand{\instMETU}{Middle East Technical University, 06531 Ankara, Turkey}
\newcommand{\instMEPhI}{Moscow Physical Engineering Institute, Moscow 115409, Russian Federation}
\newcommand{\instNagoya}{Graduate School of Science, Nagoya University, Nagoya 464-8602, Japan}
\newcommand{\instNagoyaKMI}{Kobayashi-Maskawa Institute, Nagoya University, Nagoya 464-8602, Japan}
\newcommand{\instNagoyaIAR}{Institute for Advanced Research, Nagoya University, Nagoya 464-8602, Japan}
\newcommand{\instNaraWu}{Nara Women's University, Nara 630-8506, Japan}
\newcommand{\instUNAM}{National Autonomous University of Mexico, Mexico City, Mexico}
\newcommand{\instNTUTaiwan}{Department of Physics, National Taiwan University, Taipei 10617, Taiwan}
\newcommand{\instNUUTaiwan}{National United University, Miao Li 36003, Taiwan}
\newcommand{\instKrakow}{H. Niewodniczanski Institute of Nuclear Physics, Krakow 31-342, Poland}
\newcommand{\instNiigata}{Niigata University, Niigata 950-2181, Japan}
\newcommand{\instNSU}{Novosibirsk State University, Novosibirsk 630090, Russian Federation}
\newcommand{\instOkinawa}{Okinawa Institute of Science and Technology, Okinawa 904-0495, Japan}
\newcommand{\instOsakaCity}{Osaka City University, Osaka 558-8585, Japan}
\newcommand{\instRCNP}{Research Center for Nuclear Physics, Osaka University, Osaka 567-0047, Japan}
\newcommand{\instPNNL}{Pacific Northwest National Laboratory, Richland, Washington 99352, U.S.A.}
\newcommand{\instPanjab}{Panjab University, Chandigarh 160014, India}
\newcommand{\instPeking}{Peking University, Beijing 100871, China}
\newcommand{\instPanjabPAU}{Punjab Agricultural University, Ludhiana 141004, India}
\newcommand{\instRIKENMSL}{Meson Science Laboratory, Cluster for Pioneering Research, RIKEN, Saitama 351-0198, Japan}
\newcommand{\instRIKEN}{Theoretical Research Division, Nishina Center, RIKEN, Saitama 351-0198, Japan}
\newcommand{\instXavier}{St. Francis Xavier University, Antigonish, Nova Scotia, B2G 2W5, Canada}
\newcommand{\instSeoul}{Seoul National University, Seoul 08826, South Korea}
\newcommand{\instShandong}{Shandong University, Jinan 250100, China}
\newcommand{\instSPU}{Showa Pharmaceutical University, Tokyo 194-8543, Japan}
\newcommand{\instSoochow}{Soochow University, Suzhou 215006, China}
\newcommand{\instSoongsil}{Soongsil University, Seoul 06978, South Korea}
\newcommand{\instLjubljanaJSI}{J. Stefan Institute, 1000 Ljubljana, Slovenia}
\newcommand{\instKyiv}{Taras Shevchenko National Univ. of Kiev, Kiev, Ukraine}
\newcommand{\instTata}{Tata Institute of Fundamental Research, Mumbai 400005, India}
\newcommand{\instTUM}{Department of Physics, Technische Universit\"{a}t M\"{u}nchen, 85748 Garching, Germany}
\newcommand{\instECUTUM}{Excellence Cluster Universe, Technische Universit\"{a}t M\"{u}nchen, 85748 Garching, Germany}
\newcommand{\instTelAviv}{Tel Aviv University, School of Physics and Astronomy, Tel Aviv, 69978, Israel}
\newcommand{\instToho}{Toho University, Funabashi 274-8510, Japan}
\newcommand{\instTohoku}{Department of Physics, Tohoku University, Sendai 980-8578, Japan}
\newcommand{\instTitech}{Tokyo Institute of Technology, Tokyo 152-8550, Japan}
\newcommand{\instTokyoMetropolitan}{Tokyo Metropolitan University, Tokyo 192-0397, Japan}
\newcommand{\instUAS}{Universidad Autonoma de Sinaloa, Sinaloa 80000, Mexico}
\newcommand{\instNapoliUNIV}{Dipartimento di Scienze Fisiche, Universit\`{a} di Napoli Federico II, I-80126 Napoli, Italy}
\newcommand{\instNapoliUNIVA}{Dipartimento di Agraria, Universit\`{a} di Napoli Federico II, I-80055 Portici (NA), Italy}
\newcommand{\instPadovaUNIV}{Dipartimento di Fisica e Astronomia, Universit\`{a} di Padova, I-35131 Padova, Italy}
\newcommand{\instPerugiaUNIV}{Dipartimento di Fisica, Universit\`{a} di Perugia, I-06123 Perugia, Italy}
\newcommand{\instPisaUNIV}{Dipartimento di Fisica, Universit\`{a} di Pisa, I-56127 Pisa, Italy}
\newcommand{\instRomaUNIV}{Universit\`{a} di Roma ``La Sapienza,'' I-00185 Roma, Italy}
\newcommand{\instRomaTreUNIV}{Dipartimento di Matematica e Fisica, Universit\`{a} di Roma Tre, I-00146 Roma, Italy}
\newcommand{\instTorinoUNIV}{Dipartimento di Fisica, Universit\`{a} di Torino, I-10125 Torino, Italy}
\newcommand{\instTriesteUNIV}{Dipartimento di Fisica, Universit\`{a} di Trieste, I-34127 Trieste, Italy}
\newcommand{\instMontreal}{Universit\'{e} de Montr\'{e}al, Physique des Particules, Montr\'{e}al, Qu\'{e}bec, H3C 3J7, Canada}
\newcommand{\instIJCLab}{Universit\'{e} Paris-Saclay, CNRS/IN2P3, IJCLab, 91405 Orsay, France}
\newcommand{\instIPHC}{Universit\'{e} de Strasbourg, CNRS, IPHC, UMR 7178, 67037 Strasbourg, France}
\newcommand{\instAdelaide}{Department of Physics, University of Adelaide, Adelaide, South Australia 5005, Australia}
\newcommand{\instBonn}{University of Bonn, 53115 Bonn, Germany}
\newcommand{\instUBC}{University of British Columbia, Vancouver, British Columbia, V6T 1Z1, Canada}
\newcommand{\instCincinnati}{University of Cincinnati, Cincinnati, Ohio 45221, U.S.A.}
\newcommand{\instFlorida}{University of Florida, Gainesville, Florida 32611, U.S.A.}
\newcommand{\instHamburg}{University of Hamburg, 20148 Hamburg, Germany}
\newcommand{\instHawaii}{University of Hawaii, Honolulu, Hawaii 96822, U.S.A.}
\newcommand{\instHeidelberg}{University of Heidelberg, 68131 Mannheim, Germany}
\newcommand{\instLjubljanaUniLJ}{Faculty of Mathematics and Physics, University of Ljubljana, 1000 Ljubljana, Slovenia}
\newcommand{\instLouisville}{University of Louisville, Louisville, Kentucky 40292, U.S.A.}
\newcommand{\instMalaya}{National Centre for Particle Physics, University Malaya, 50603 Kuala Lumpur, Malaysia}
\newcommand{\instLjubljanaUM}{University of Maribor, 2000 Maribor, Slovenia}
\newcommand{\instMelbourne}{School of Physics, University of Melbourne, Victoria 3010, Australia}
\newcommand{\instMississippi}{University of Mississippi, University, Mississippi 38677, U.S.A.}
\newcommand{\instUOM}{University of Miyazaki, Miyazaki 889-2192, Japan}
\newcommand{\instNovaGorica}{University of Nova Gorica, 5000 Nova Gorica, Slovenia}
\newcommand{\instPittsburgh}{University of Pittsburgh, Pittsburgh, Pennsylvania 15260, U.S.A.}
\newcommand{\instUSTC}{University of Science and Technology of China, Hefei 230026, China}
\newcommand{\instSAlabama}{University of South Alabama, Mobile, Alabama 36688, U.S.A.}
\newcommand{\instSCarolina}{University of South Carolina, Columbia, South Carolina 29208, U.S.A.}
\newcommand{\instSydney}{School of Physics, University of Sydney, New South Wales 2006, Australia}
\newcommand{\instTabuk}{Department of Physics, Faculty of Science, University of Tabuk, Tabuk 71451, Saudi Arabia}
\newcommand{\instUTokyo}{Department of Physics, University of Tokyo, Tokyo 113-0033, Japan}
\newcommand{\instIPMU}{Kavli Institute for the Physics and Mathematics of the Universe (WPI), University of Tokyo, Kashiwa 277-8583, Japan}
\newcommand{\instVictoria}{University of Victoria, Victoria, British Columbia, V8W 3P6, Canada}
\newcommand{\instVPI}{Virginia Polytechnic Institute and State University, Blacksburg, Virginia 24061, U.S.A.}
\newcommand{\instWayneState}{Wayne State University, Detroit, Michigan 48202, U.S.A.}
\newcommand{\instYamagata}{Yamagata University, Yamagata 990-8560, Japan}
\newcommand{\instYerevan}{Alikhanyan National Science Laboratory, Yerevan 0036, Armenia}
\newcommand{\instYonsei}{Yonsei University, Seoul 03722, South Korea}
\affiliation{\instCPPM}
\affiliation{\instBeihang}
\affiliation{\instBNL}
\affiliation{\instBINP}
\affiliation{\instCMU}
\affiliation{\instCinvestavIPN}
\affiliation{\instPrague}
\affiliation{\instChiangMai}
\affiliation{\instChiba}
\affiliation{\instChonnam}
\affiliation{\instConacyt}
\affiliation{\instDESY}
\affiliation{\instDuke}
\affiliation{\instITAR}
\affiliation{\instEri}
\affiliation{\instJuelich}
\affiliation{\instFuJen}
\affiliation{\instFudan}
\affiliation{\instGoettingen}
\affiliation{\instGifu}
\affiliation{\instSOKENDAI}
\affiliation{\instGyeongsang}
\affiliation{\instHanyang}
\affiliation{\instKEK}
\affiliation{\instJPARC}
\affiliation{\instHSE}
\affiliation{\instIISER}
\affiliation{\instIITBhubaneswar}
\affiliation{\instIITGuwahati}
\affiliation{\instIITHyderabad}
\affiliation{\instIITMadras}
\affiliation{\instIndiana}
\affiliation{\instIHEPRussia}
\affiliation{\instHEPHYVienna}
\affiliation{\instIHEPChina}
\affiliation{\instIPP}
\affiliation{\instIOP}
\affiliation{\instIFIC}
\affiliation{\instFrascati}
\affiliation{\instNapoliINFN}
\affiliation{\instPadovaINFN}
\affiliation{\instPerugiaINFN}
\affiliation{\instPisaINFN}
\affiliation{\instRomaINFN}
\affiliation{\instRomaTreINFN}
\affiliation{\instTorinoINFN}
\affiliation{\instTriesteINFN}
\affiliation{\instJAEA}
\affiliation{\instMainz}
\affiliation{\instGiessen}
\affiliation{\instKarlsruhe}
\affiliation{\instKitasato}
\affiliation{\instKISTI}
\affiliation{\instKorea}
\affiliation{\instKSU}
\affiliation{\instKyungpook}
\affiliation{\instLPI}
\affiliation{\instLNNU}
\affiliation{\instLMU}
\affiliation{\instLuther}
\affiliation{\instMNITJaipur}
\affiliation{\instMPP}
\affiliation{\instMPGHLL}
\affiliation{\instMcGill}
\affiliation{\instMEPhI}
\affiliation{\instNagoya}
\affiliation{\instNagoyaKMI}
\affiliation{\instNagoyaIAR}
\affiliation{\instNaraWu}
\affiliation{\instNTUTaiwan}
\affiliation{\instNUUTaiwan}
\affiliation{\instKrakow}
\affiliation{\instNiigata}
\affiliation{\instNSU}
\affiliation{\instOkinawa}
\affiliation{\instOsakaCity}
\affiliation{\instRCNP}
\affiliation{\instPNNL}
\affiliation{\instPanjab}
\affiliation{\instPeking}
\affiliation{\instPanjabPAU}
\affiliation{\instRIKENMSL}
\affiliation{\instSeoul}
\affiliation{\instSPU}
\affiliation{\instSoochow}
\affiliation{\instSoongsil}
\affiliation{\instLjubljanaJSI}
\affiliation{\instKyiv}
\affiliation{\instTata}
\affiliation{\instTUM}
\affiliation{\instTelAviv}
\affiliation{\instToho}
\affiliation{\instTohoku}
\affiliation{\instTitech}
\affiliation{\instTokyoMetropolitan}
\affiliation{\instUAS}
\affiliation{\instNapoliUNIV}
\affiliation{\instPadovaUNIV}
\affiliation{\instPerugiaUNIV}
\affiliation{\instPisaUNIV}
\affiliation{\instRomaUNIV}
\affiliation{\instRomaTreUNIV}
\affiliation{\instTorinoUNIV}
\affiliation{\instTriesteUNIV}
\affiliation{\instMontreal}
\affiliation{\instIJCLab}
\affiliation{\instIPHC}
\affiliation{\instAdelaide}
\affiliation{\instBonn}
\affiliation{\instUBC}
\affiliation{\instCincinnati}
\affiliation{\instFlorida}
\affiliation{\instHawaii}
\affiliation{\instHeidelberg}
\affiliation{\instLjubljanaUniLJ}
\affiliation{\instLouisville}
\affiliation{\instMalaya}
\affiliation{\instLjubljanaUM}
\affiliation{\instMelbourne}
\affiliation{\instMississippi}
\affiliation{\instUOM}
\affiliation{\instPittsburgh}
\affiliation{\instUSTC}
\affiliation{\instSAlabama}
\affiliation{\instSCarolina}
\affiliation{\instSydney}
\affiliation{\instUTokyo}
\affiliation{\instIPMU}
\affiliation{\instVictoria}
\affiliation{\instVPI}
\affiliation{\instWayneState}
\affiliation{\instYamagata}
\affiliation{\instYerevan}
\affiliation{\instYonsei}
  \author{F.~Abudin{\'e}n}\affiliation{\instTriesteINFN} 
  \author{I.~Adachi}\affiliation{\instKEK}\affiliation{\instSOKENDAI} 
  \author{R.~Adak}\affiliation{\instFudan} 
  \author{K.~Adamczyk}\affiliation{\instKrakow} 
  \author{P.~Ahlburg}\affiliation{\instBonn} 
  \author{J.~K.~Ahn}\affiliation{\instKorea} 
  \author{H.~Aihara}\affiliation{\instUTokyo} 
  \author{N.~Akopov}\affiliation{\instYerevan} 
  \author{A.~Aloisio}\affiliation{\instNapoliUNIV}\affiliation{\instNapoliINFN} 
  \author{F.~Ameli}\affiliation{\instRomaINFN} 
% ...
\collaboration{Belle II Collaboration}

%% file: abstract.tex
The reconstruction of tau-pair production, $\epem \to \tau^{+}\tau^{-}$, from the subsequent 3-prong  ($\tau^{+} \rightarrow \pi^{+} \pi^{-} \pi^{+} \bar{\nu}_{\tau}$) and 1-prong ($\tau^{-} \to \ell^{-} \bar{\nu}_{\ell} \nu_{\tau}$, $\tau^{-} \to h^{-} \nu_{\tau}$ or $\tau^{-} \to \pi^{-} \pi^0 \nu_{\tau}$) decays, is presented using 8.8~\invfb of $\epem $ collision data of {\belletwo} at 
the center-of-mass energy $\sqrt{s} = m_{\Upsilon(4S)}$. The pseudomass technique developed by the ARGUS experiment is used to measure the $\tau$-lepton mass $m_{\tau}$ in the 3-prong $\tau^{+} \to \pi^{+} \pi^{-} \pi^{+} \bar{\nu}_{\tau} $ decay, resulting in $m_{\tau} = 1777.28  \pm 0.75~{\rm (stat.)} \pm 0.33 ~{\rm (sys.)}\mevcc$. 

%% file: main.tex
\section{Introduction}
\label{sec:int}
Precise measurements of the lepton properties provide stringent tests of the Standard Model (SM) and accurate determinations of its parameters. 
For example, the SM predicts unambiguous and simple relationships among the lepton lifetime, mass, and leptonic branching fractions. 
Therefore, their experimental determination to the highest possible precision is essential; deviations from the predictions at any level could signal the presence of physics beyond our present understanding. 

\begin{figure}[hb!]
	\begin{center}
	\includegraphics[width=10cm]{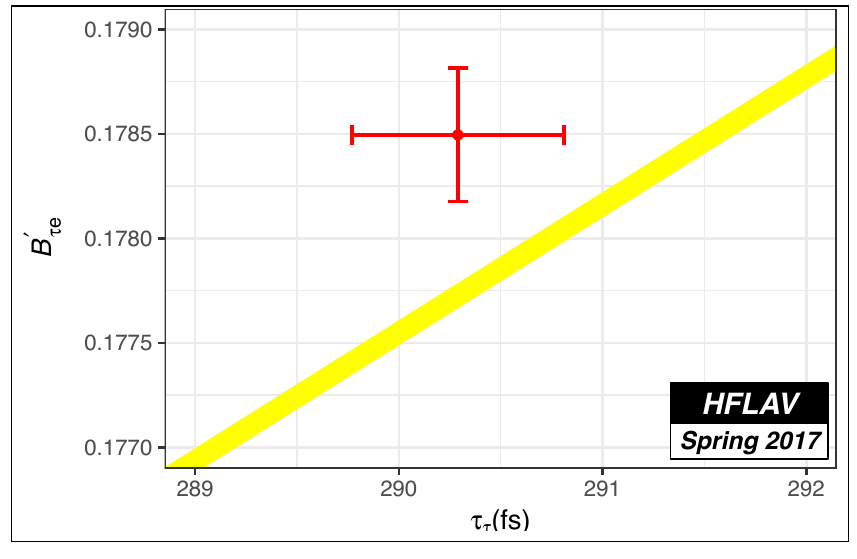}
	\caption{ Test of the SM prediction of the relation  between the $\tau$ leptonic branching fractions and the $\tau$ lifetime and mass. $B^\prime_{\tau e}$ denotes the statistical average of $B_{\tau e} = B(\tau^-\to e^-\bar{\nu}_e\nu_\tau)$ and the $B_{\tau e}$ SM prediction from the $B_{\tau \mu}$ measurement $B_{\tau e} = B_{\tau\mu} \cdot ( f_{\tau e}/ f_{\tau \mu})$. The yellow band represents the uncertainty from the $\tau$ lifetime. The plot is from Ref.~\cite{Lusiani:2018zvr}.}
\label{fig:uni}
\end{center}
\end{figure}

Figure~\ref{fig:uni} illustrates the test of the SM prediction of the relation among the $\tau$ leptonic branching fractions, $ B_{\tau\ell} = B(\tau^- \to \ell^- \bar{\nu}_\ell \nu_\tau$) with $\ell^- = e^- \; {\rm or} \; \mu^-$; the $\tau$ lifetime $\tau_\tau$; the $\tau$ mass $m_\tau$; and the respective muon parameters:
\begin{equation}
    {B}^{SM}_{\tau \ell}  =  {B}_{\mu e} \frac{\tau_\tau}{\tau_\mu} \;
    \frac{m^5_\tau}{m^5_\mu} \;
    \frac{f_{\tau\ell}}{f_{\mu e}} \;
    \frac{r_W^\tau r_\gamma^\tau}{r_W^\mu r_\gamma^\mu} \; \; \; .
\end{equation}
Here $f_{\ell \ell^\prime} = f\Big(\frac{m^2_{\ell^\prime}}{m^2_\ell}\Big) \equiv f(x) = 1 - 8x + 8x^2 - x^4 - 12x^2\ln{x}$, 
with $r_W$ and $r_\gamma$ being the weak and electromagnetic radiative corrections.
This relation is very sensitive to the value of the $\tau$ mass.

The $\tau$ leptonic branching fractions and the $\tau$ lifetime are known up to a relative precision of $~0.2\%$, far from the impressive accuracy achieved for the $\mu$ lifetime. Comparing the global average of the $e$, $\mu$ and $\tau$ masses reported by the PDG~\cite{PDG2018},
\begin{eqnarray}
m_e  & = & (0.5109989461 \pm 0.0000000031)~\mevcc, \\
m_\mu & = & (105.6583745 \pm 0.0000024)~\mevcc,   \\
m_\tau  & = & (1776.86 \pm 0.12)~\mevcc,
\end{eqnarray}
it is clear that the value of $m_\tau$ is also far less precise. The average of $m_\tau$ is dominated by the result from the BES~III experiment~\cite{BESIIImass} that makes an energy scan of $\sigma(\epem \to \tau^{+} \tau^{-})$ around the $\tau^+\tau^-$ production threshold, followed by the measurements reported by the \belle~\cite{BelleMass} and BaBar~\cite{BabarMass} collaborations in which the $\tau$ pseudomass spectrum from $\tau^- \to \pi^-\pi^+\pi^-\nu_\tau$ decay is used applying the ARGUS method~\cite{Albrecht:1992}. Though less precise, the pseudomass endpoint method 
allows us to test CPT conservation by measuring the masses of $\tau^-$ and $\tau^+$ individually. 

In this work, preliminary results on the reconstruction of $\tau$-pair production, $\epem \to \tau^{+}\tau^{-}$, from the subsequent 3-prong ($\tau^+ \to \pi^+\pi^-\pi^+ \bar{\nu}_\tau$) and 1-prong ($\tau^- \to \ell^- {\bar \nu}_\ell \nu_\tau $, $\tau^- \to h^- \nu_\tau$ or $\tau^- \to \pi^- \pi^0 \nu_\tau$) decays\footnote{Charge-conjugate modes are implied in this study.}, and the $\tau$ lepton mass measurement are presented. 

\section{Event selection}
\label{sec:candidateEventSelection}
The analysis presented here is based on 8.8\invfb of data accumulated during 2019 at the $\Upsilon(4S)$ resonance ($\sqrt{s} = 10.58$~\gev) with the {\belletwo} detector at the SuperKEKB asymmetric-energy $e^{+}e^{-}$ collider~\cite{Akai:2018mbz}.
The {\belletwo} detector consists of several subdetectors arranged around the beam pipe in a cylindrical structure. 
A superconducting solenoid, situated outside of the calorimeter, provides a 1.5 T magnetic field. Subdetectors relevant to this analysis are briefly described here; a description of the full detector is given in Ref.~\cite{Abe:2010gxa}. The innermost subdetector is the vertex detector (VXD), which includes two layers of silicon pixels and four outer layers of silicon strips. Charged-particle tracking is done using the VXD and a large helium-based small-cell central drift chamber (CDC). 
The electromagnetic calorimeter (ECL) consists of a barrel and two endcaps made of CsI(Tl) crystals. 
The $z$-axis of the laboratory frame is along the detector solenoidal axis in the direction of the electron beam. 
Events are selected by the hardware trigger; no further software trigger selection is applied. 

The Belle II experiment records data in a quite severe beam background environment. Due to the short lifetime of the $\tau$ lepton, its decay products are expected to originate 
close to the interaction point (IP). 
Thus, the $\tau$-pair candidate events are selected by requiring only four charged tracks, with zero net charge, originating from a narrow window close to the IP. 
The beam-induced photon background has typically low energy and can be largely reduced by imposing a cut on the minimum energy of the photons. 

The $\pi^0 \to \gamma \gamma$ candidates are reconstructed from photon candidates 
with $17\degree < \theta_{\gamma} < 150\degree$ 
and an energy threshold of 100~\mev in the invariant mass window $0.115$~GeV/$c^{2}$ $< M_{\gamma\gamma} < 0.152$~GeV/$c^{2}$. 

In the $\epem \to \tau^{+}\tau^{-}$ centre-of-mass system (CMS), both $\tau$ leptons are back-to-back and their decay products are well separated in two opposite hemispheres defined by the plane perpendicular to the thrust axis~\cite{Brandt:1964sa,Farhi:1977sg}. The thrust axis $\hat{n}_{\rm thrust}$ is defined such that the value $V_{\rm thrust}$
\begin{equation}
	V_{\rm thrust} = \sum_i \frac{|{\vec{p}_i}^{\hspace{0.1cm} \rm CMS} \cdot \hat{n}_{\rm thrust}|}{\sum |\vec{p}^{\hspace{0.1cm} \rm CMS}_i|}
\end{equation}
is maximised. Here, ${\vec{p}_i}^{\hspace{0.1cm}\rm CMS}$ is the CMS momentum of each charged particle, $\pi^0$ or photon. 
A pion mass hypothesis is used for all charged tracks. Given the vector $\hat{n}_{\rm thrust}$, one hemisphere is expected to contain the products of 3-prong $\tau$ decay, while the other one the products of 1-prong $\tau$ decay. 

The  identification of charged particles is based on $E_{\rm ECL}/P_{\rm lab}$, where $E_{\rm ECL}$ is the energy deposit in the ECL, while $P_{\rm lab}$ is the momentum of the particle 
measured in the tracking systems (VXD+CDC). For the reconstruction of the 3-prong decay, pions are required to satisfy the condition, $E_{\rm ECL}/P_{\rm lab} < 0.8$. 
This requirement enhances the selection of $\tau^+ \to \pi^+\pi^-\pi^+\bar{\nu}_\tau$ in 3-prong decays.
The $\pi^0$ are vetoed on the 3-prong side, in order to reduce the background contamination from  $\tau^+ \to \pi^+\pi^-\pi^+\pi^0\bar{\nu}_\tau$ decays. 
The 1-prong side is expected to contain one charged track and at most one neutral pion.

%

\begin{figure}[h!]
	\begin{center}
	\includegraphics[width=12cm]{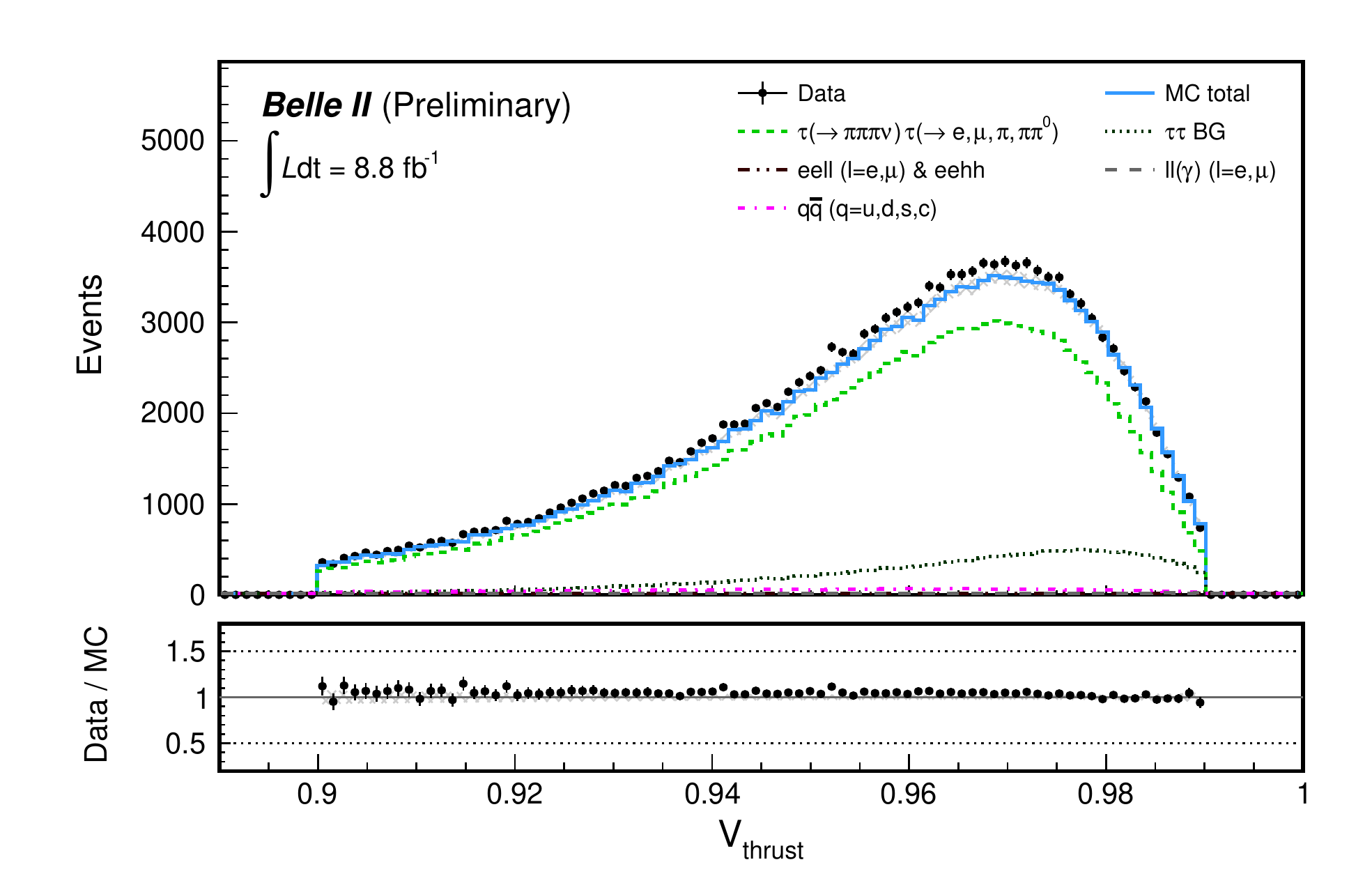} \\
    \includegraphics[width=12cm]{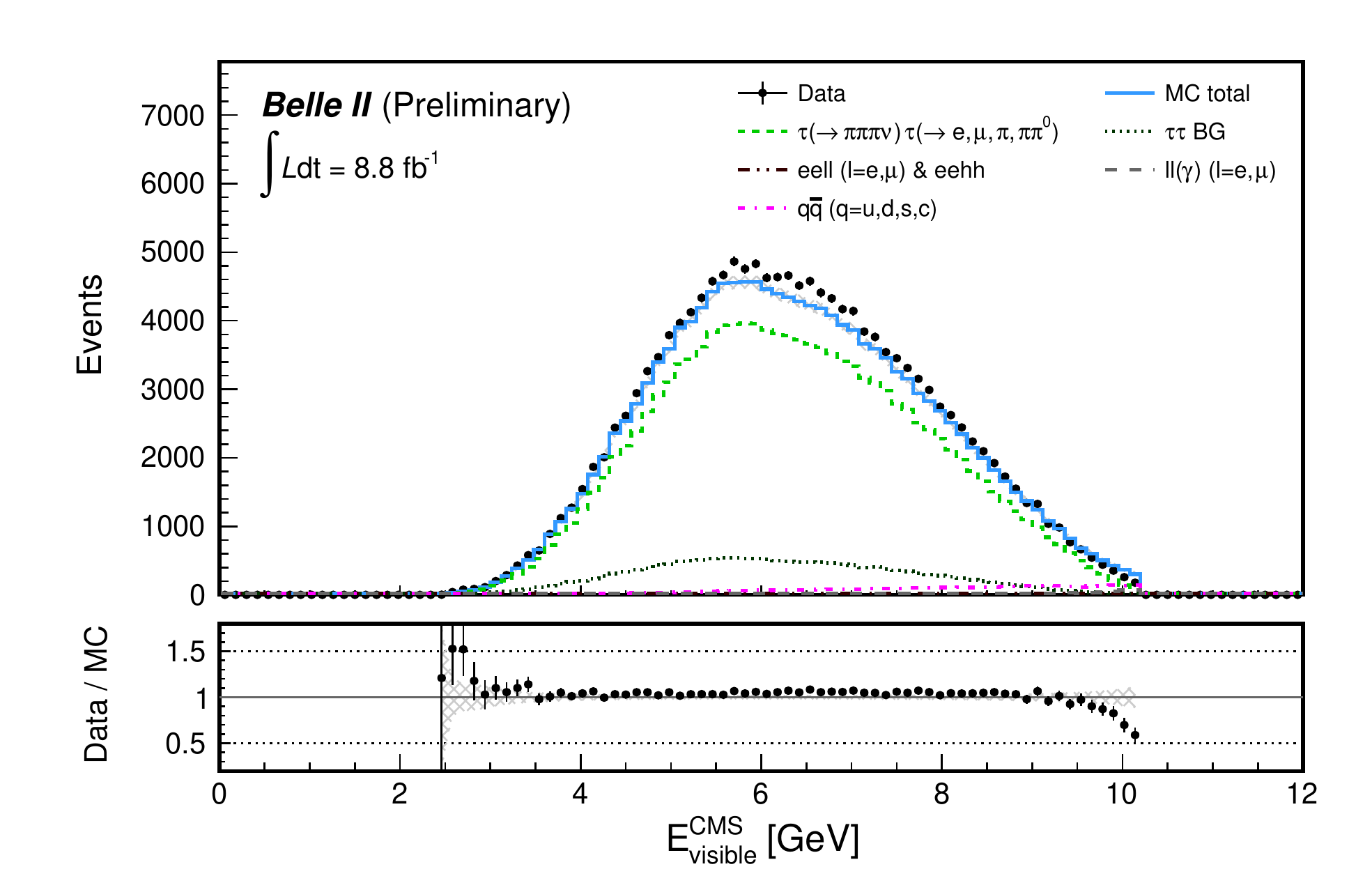}
	\caption{The distributions of $V_{\rm thrust}$ (top) and visible energy in CMS (bottom) in data and Monte Carlo after the selection and trigger correction. 
	The light green dashed curve shows the simulated distribution of the $\epem \to \tau^+\tau^-$ process with subsequent 3-prong ($\tau^+ \to \pi^+\pi^-\pi^+\bar{\nu}_\tau$) and 1-prong ($\tau^- \to \ell^- \bar{\nu_\ell} \nu_\tau$, $\tau^- \to h^- \nu_\tau$ or $\tau^- \to \pi^- \pi^0 \nu_\tau$) decays, while the dark green dotted distribution corresponds to the background contribution from other $\tau$ decay modes. 
	The background contamination from $q {\bar q}$ process is shown with dash-dotted magenta curves. 
	The sum of all Monte Carlo contributions and the corresponding total statistical uncertainties are shown by the solid blue curve and the grey hatched area, respectively. 
	The Monte Carlo distributions are rescaled to a luminosity of 8.8\invfb of data and reweighted according to the trigger efficiency measured in data. 
	The bottom panel of each plot shows the ratio of the data and total MC prediction.}
\label{fig:thrustAndVisibleEnergyCMS}
\end{center}
\end{figure}

There are processes other than the $\tau$-pair production that may satisfy selection criteria, such as 
$\epem \to \Upsilon(4S) \to B {\bar B}$, $\epem \to q {\bar q}$ (with $q = u, d, s, c$), $\epem \to \ell^{+} \ell^{-} \gamma$ (with $\ell = e, \mu$) and two-photon processes $\epem \to e^{+}e^{-}\ell^{+}\ell^{-}$ and 
$\epem \to e^{+}e^{-}q\bar{q}$. 
In order to reduce the background contamination from $\epem \to q\bar q$ processes, 
any event with a photon having $E > 200$\mev that is not the daughter of a $\pi^0$ is rejected.
To suppress other background sources, differences in the distributions of the thrust value and the total visible energy of event in the CMS are used, 
where the energy of the pions is calculated from their momentum and mass.
A signal-to-background optimization leads to $0.9\leq V_{\rm thrust} \leq 0.99$ and the visible energy in CMS, $E_{\rm visible}^{\rm CMS}$, within $[2.5,10.2]~\gev$.
After the trigger efficiency correction, the data and Monte Carlo agreement is within a few percent, as shown in Fig.~\ref{fig:thrustAndVisibleEnergyCMS}.

\begin{figure}[h!]
	\begin{center}
	\includegraphics[width=12cm]{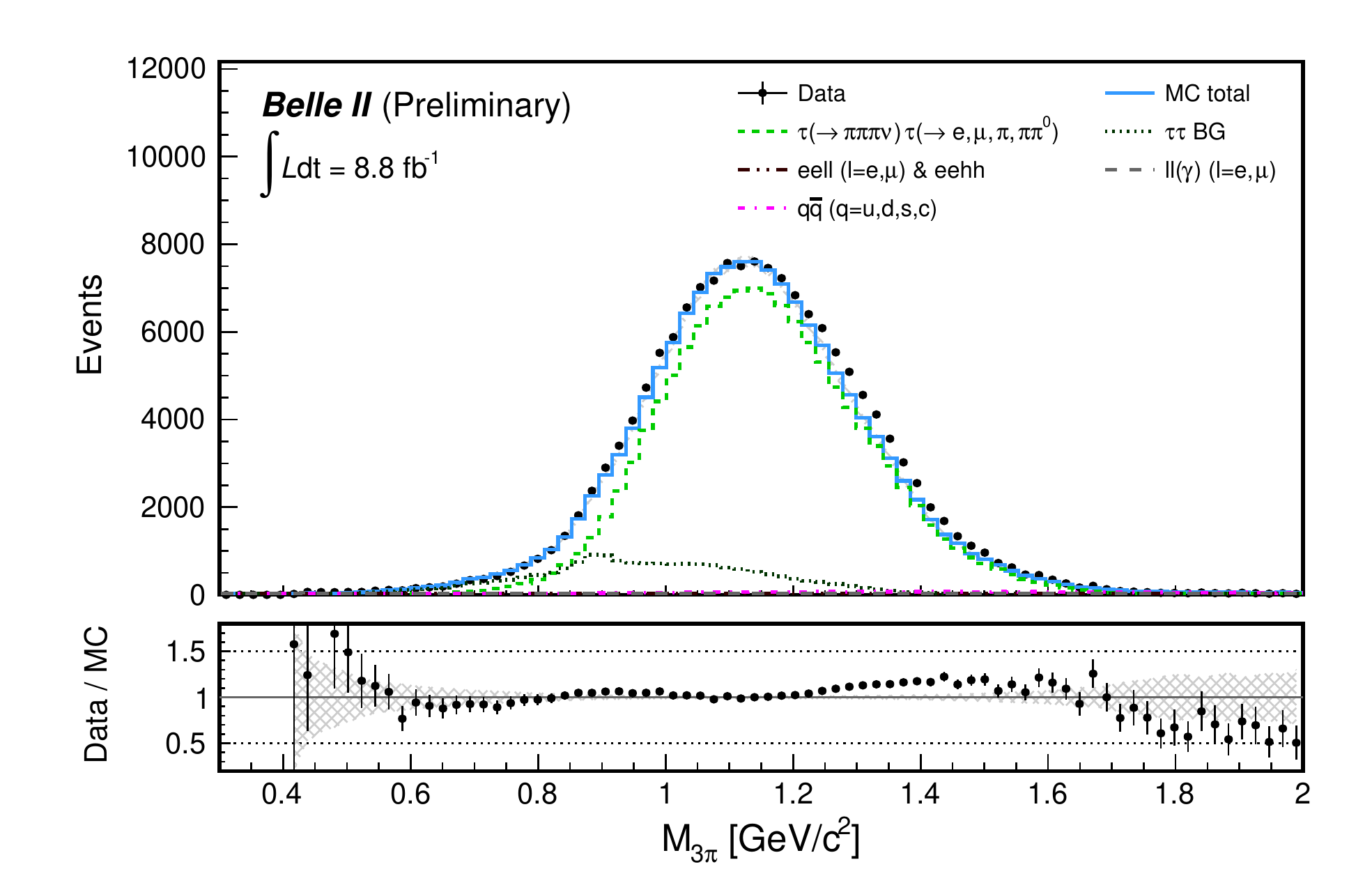} 
	\caption{The 
	invariant mass $M_{3\pi}$ of the three tracks on the 3-prong side after the selection criteria. The description of the plot is the same as that of Fig.~\ref{fig:thrustAndVisibleEnergyCMS}.}
\label{fig:invmass}
\end{center}
\end{figure}

After the selection is applied, there remain about 150k events of $\epem~\to \tau^+\tau^-$ events with the subsequent 3-prong ($\tau^+ \to \pi^+\pi^-\pi^+\bar{\nu}_\tau$) and 1-prong ($\tau^- \to \ell^- \bar{\nu}_\ell \nu_\tau$, $\tau^- \to h^- \nu_\tau$ or $\tau^- \to \pi^- \pi^0 \nu_\tau$) decays. The invariant mass of the three tracks on the 3-prong side ($M_{3\pi}$) is shown in Fig.~\ref{fig:invmass}. The efficiency of reconstructing the signal events is $16.6\%$; the purity of the sample $84.5 \%$. The background contamination in the $M_{3\pi}$ distribution arises mainly from $\tau$ decays other than $\tau^+ \to \pi^+ \pi^- \pi^+ \bar{\nu}_\tau$. The largest contribution is from $\tau^+ \to \pi^+ \pi^- \pi^+ \pi^0\bar{\nu}_\tau$ decays where the $\pi^0$ is not reconstructed, followed by the contribution from events with $K-\pi$ misidentification. This is due to the fact that no pion identification is used in the analysis. 

\section{Measurement of the $\tau$ lepton mass}
\label{sec:measurement}
\begin{figure}[h]
	\begin{center}
        \includegraphics[width=12cm]{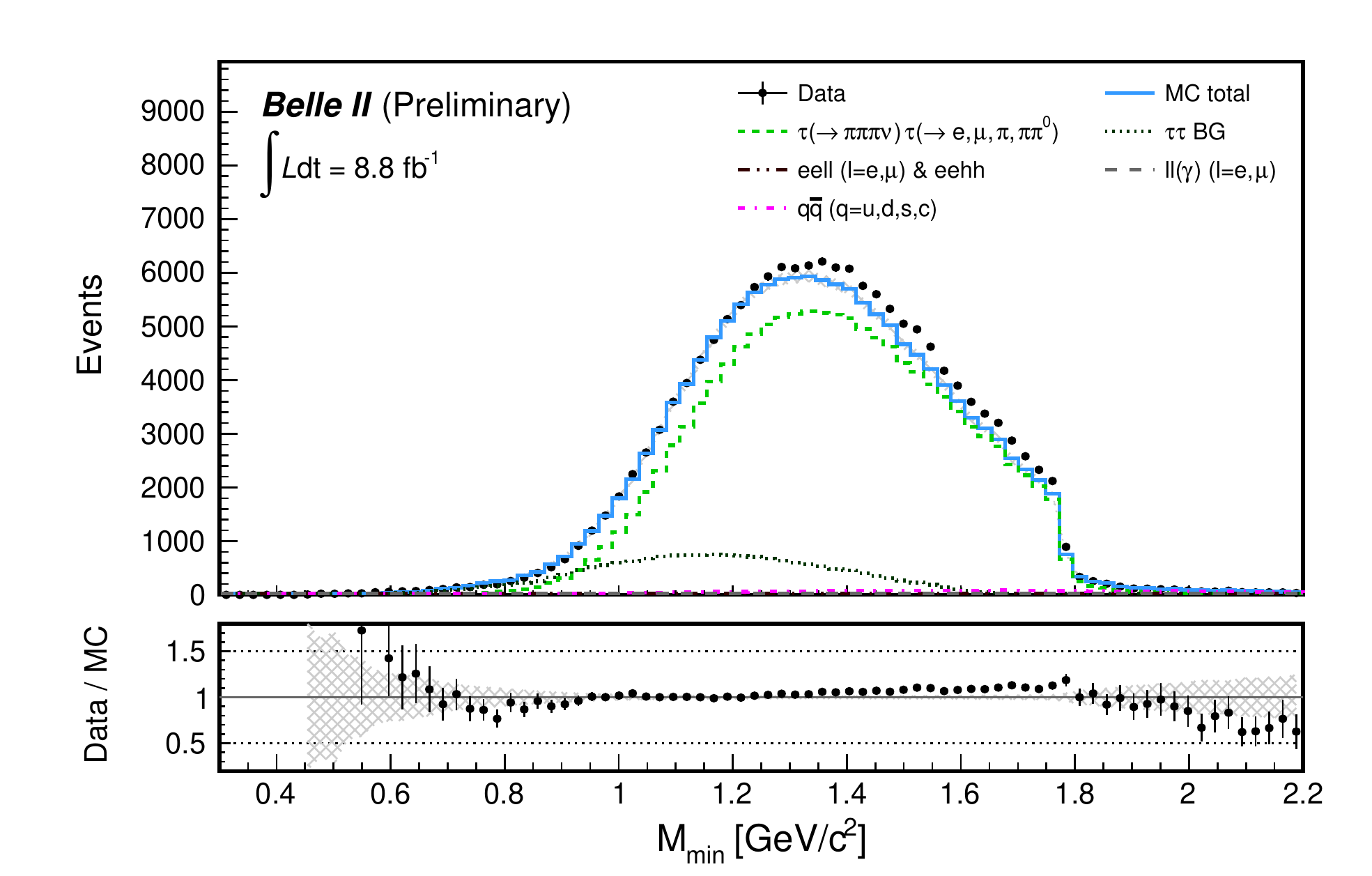}\\
        \includegraphics[width=12cm]{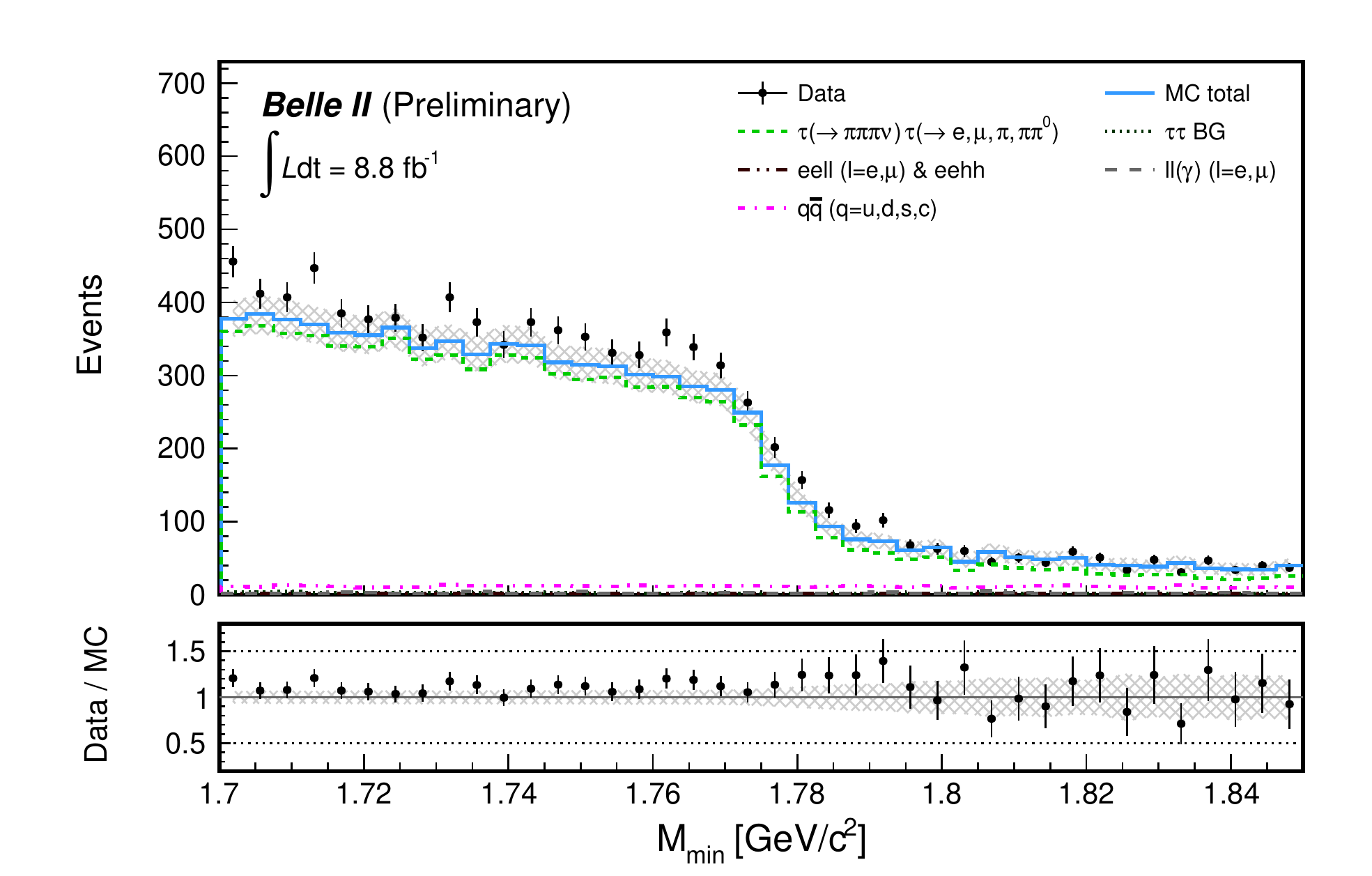}
		\caption{Distribution of the pseudomass for $\epem \to \tau^+  \tau^- $ process with subsequent 3-prong ($\tau^+ \to \pi^+\pi^-\pi^+\bar{\nu}_\tau$) and 1-prong ($\tau^- \to \ell^- \bar{\nu_\ell} \nu_\tau$, $\tau^- \to h^- \nu_\tau$ or $\tau^- \to \pi^- \pi^0 \nu_\tau$) decays in the entire range (up), and in the range $1.70$ to $1.85$~\gevcc (bottom). The description of the plot is the same as that of Fig.~\ref{fig:thrustAndVisibleEnergyCMS}.}	
	\label{fig:pseudomasstrig}
	\end{center}
\end{figure}

The analysis procedure is blinded while establishing the technique, selection criteria and evaluation of the systematic uncertainties. 
The $\tau$-lepton mass measurement is performed following the pseudomass method developed by the ARGUS collaboration \cite{Albrecht:1992}. The pseudomass is defined as
\begin{equation}
	M_{\rm min} = \sqrt{M_{3\pi}^2 + 2(E_{\rm beam} - E_{3\pi})(E_{3\pi} - P_{3\pi})} \leq m_\tau,
	\label{eq:mmin}
\end{equation}
where $E_{\rm beam}$ is the energy of one of the beams in CMS, and $M_{3\pi}$, $E_{3\pi}$, $P_{3\pi}$ stand for the invariant mass, energy and momentum, respectively, of the three pion system in CMS. In the absence of initial (ISR) and final (FSR) state radiations, and a perfect measurement of the four-momentum of the hadronic system, the distribution of $M_{\rm min}$ extends up to and has a sharp edge at the mass of the $\tau$ lepton, $m_\tau$. The ISR/FSR and the detector resolution smear the endpoint and result in a large tail in the pseudomass distribution, as seen on the left panel of Fig.~\ref{fig:pseudomasstrig}. The mass of the $\tau$ lepton $m_{\tau}$ is then measured by determining the position of the endpoint. 

The $M_{\rm min}$ distribution from $1.70$\gevcc to $1.85$\gevcc is of particular interest for extracting the $\tau$ mass. 
Unlike the invariant mass distribution where the main background contamination is due to misidentified $\tau^+ \to \pi^+\pi^-\pi^+ \bar{\nu}_\tau$ decays, in the region of interest of the pseudomass distribution all background contributions are negligible except for $e^+e^- \to q\bar{q}$. While $\tau^+ \to \pi^+\pi^-\pi^+ \bar{\nu}_\tau$ decays show the endpoint behaviour, the background processes in the selected region have a featureless $M_{\rm min}$ distribution.

An empirical probability density function (p.d.f.)
\begin{equation}
	F(M,\vec{P}) = (P_3 + P_4 M)\cdot \tan^{-1}[(M - P_1)/P_2] + P_5 M + 1 \;,
	\label{eq:pdf}
\end{equation}
is used to estimate the $\tau$ lepton mass from the $\epem \to (\tau^+\to \pi^+\pi^-\pi^+\bar{\nu}_\tau)(\tau^-\to e^-, \mu^-, \pi^-, \pi^-\pi^0)$ Monte Carlo sample~\cite{Jadach:1999vf, Jadach:1990mz}, in which the parameter $P_1$ is an estimator of the $\tau$ lepton mass. The fit results in 
\begin{equation}
P_1 = \MCTauMassFit \mevcc\; .
\label{eq:P1}
\end{equation}
At the generator level, $m_{\tau}$ is set to 1777\mevcc; thus, the $P_1$ parameter shows a bias in the estimation of the $\tau$ mass as observed in previous measurements of the $\tau$ mass using the pseudomass method~\cite{BelleMass, BabarMass, Albrecht:1992}. 
The average bias of $P_1$ is estimated to be \MCEstimatorBias\mevcc independently of the generated $\tau$ mass, by performing fits to Monte Carlo samples that were generated using $m_{\tau}$ values shifted with respect to the nominal values.

\section{Systematic Uncertainties}
\label{sec:sys}
\input{systematics}

\section{Results}
\label{sec:results}
\begin{figure}[h]
	\begin{center}
		\includegraphics[width=12cm]{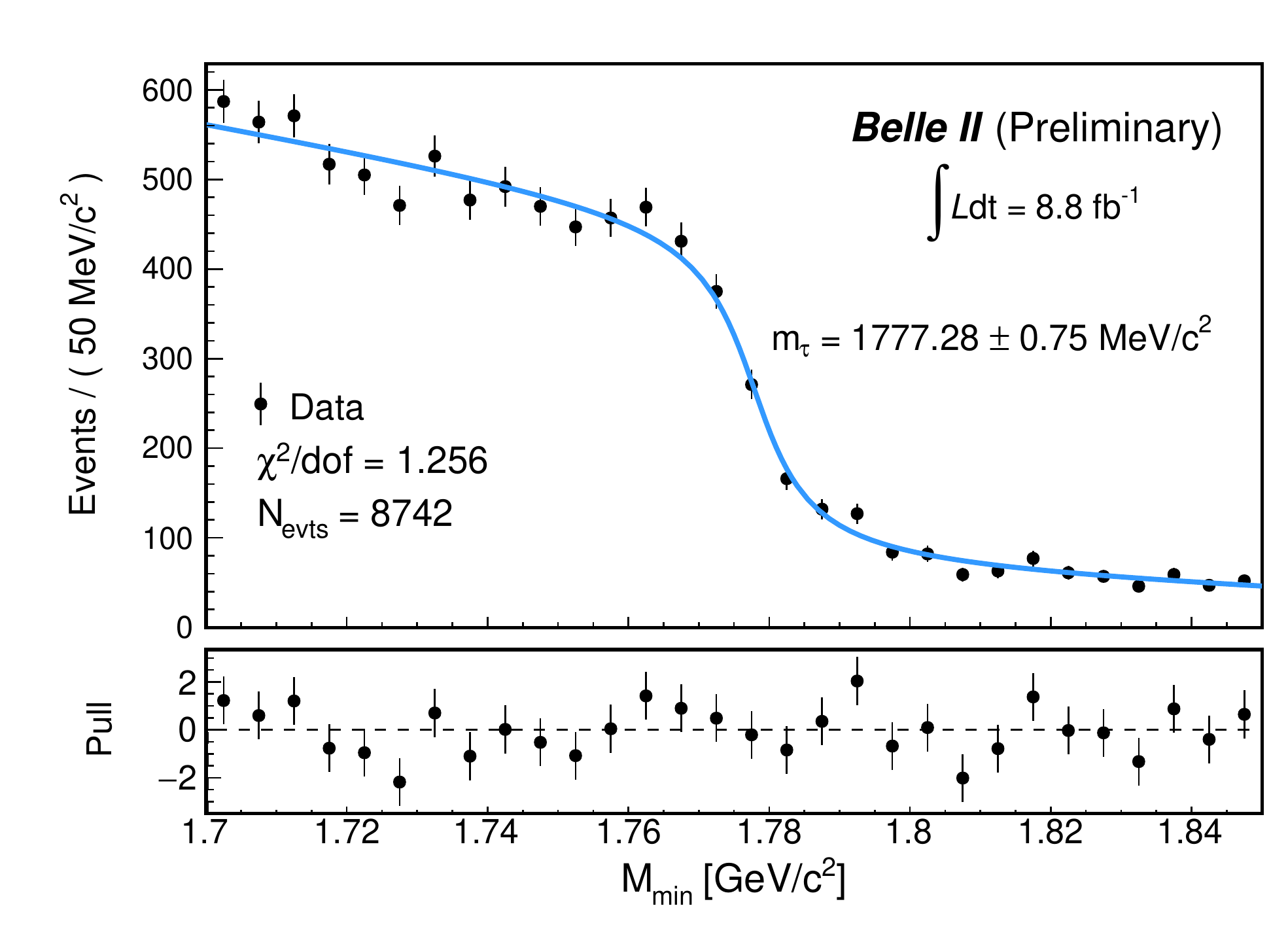} 
		\caption{The pseudomass ($M_{\rm min}$) distribution in the data sample (black points) and the results of the fit (blue line). }
		\label{fig:fit_data}
	\end{center}
\end{figure}

After unblinding the data, the fit procedure is performed on the data.  
Figure~\ref{fig:fit_data} shows the result of the fit, indicating a $P_{1}$ estimator value of $1778.00\pm0.75\mevcc$. Using the correction factor obtained in section~\ref{sec:measurement}, and the systematic uncertainties described in section~\ref{sec:sys}, the mass of the $\tau$ lepton is measured as:
\begin{equation}
    m_{\tau} = 1777.28  \pm 0.75~{\rm (stat.)} \pm 0.33 ~{\rm (sys.)}~\mevcc
    \label{eq:mtau}
\end{equation}

\section{Conclusions}

\begin{figure}[h!]
    \centering
    \includegraphics[width=12cm]{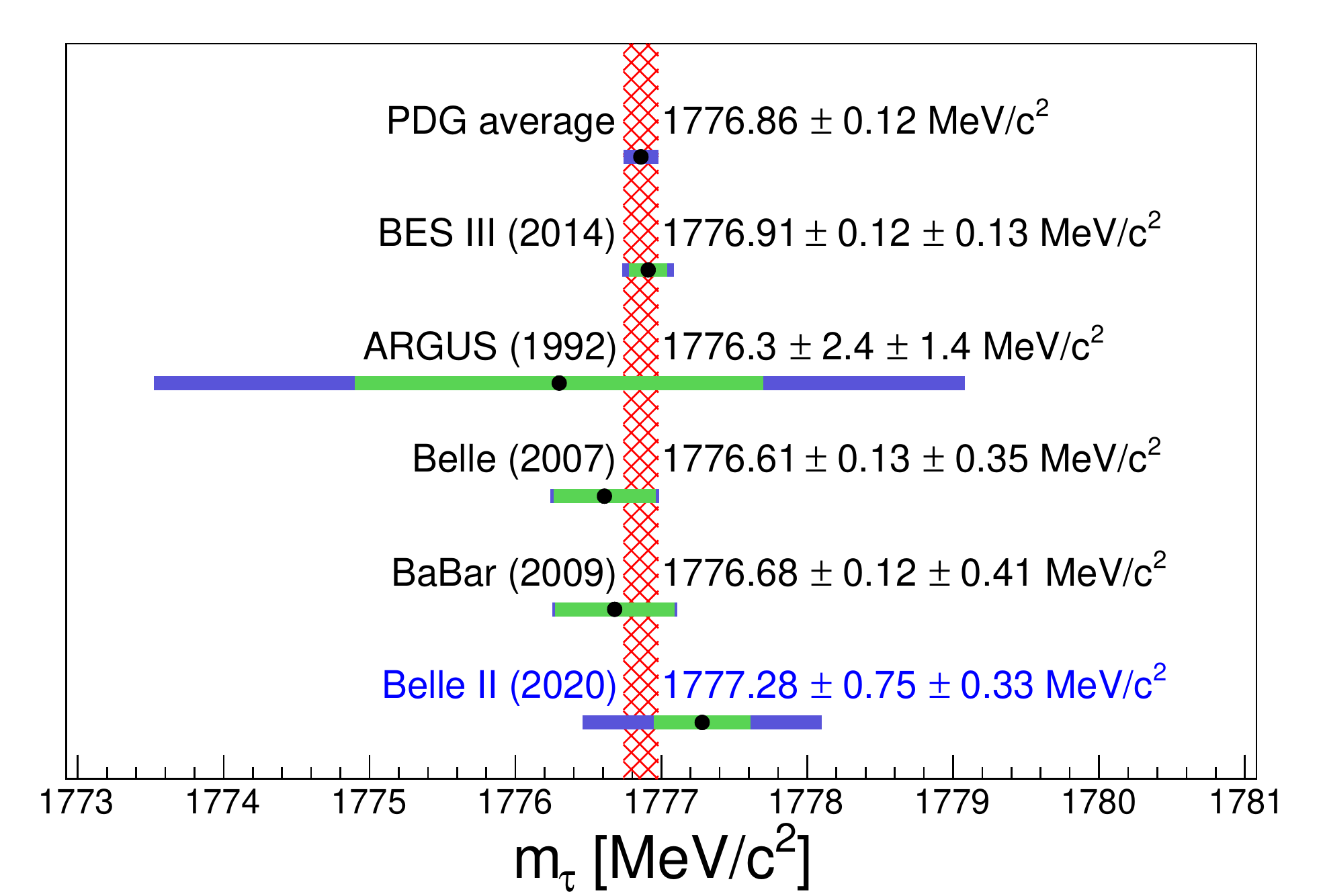}
    \caption{The comparison of the $\tau$ mass measurements obtained in this analysis (in blue text) with the PDG average and measurements from various experiments. The green and blue bands indicate the systematic and total uncertainties, respectively.}
    \label{fig:taumassworld}
\end{figure}

In summary, the mass of the $\tau$ lepton has been measured using the pseudomass method in a blinded analysis procedure. Using the $\tau^+ \to \pi^+\pi^-\pi^+\bar{\nu}_\tau$ decays in 8.8\invfb of data, the mass of the $\tau$ lepton has been found to be $m_{\tau} = 1777.28 \pm 0.75~{\rm (stat.)} \pm 0.33 ~{\rm (sys.)}\mevcc$. This measurement is in good agreement with the current world average~\cite{PDG2018}. Figure~\ref{fig:taumassworld} shows this result compared to other recent $\tau$ mass measurements.

\begin{figure}[h!]
    \centering
    \includegraphics[width=0.7\textwidth]{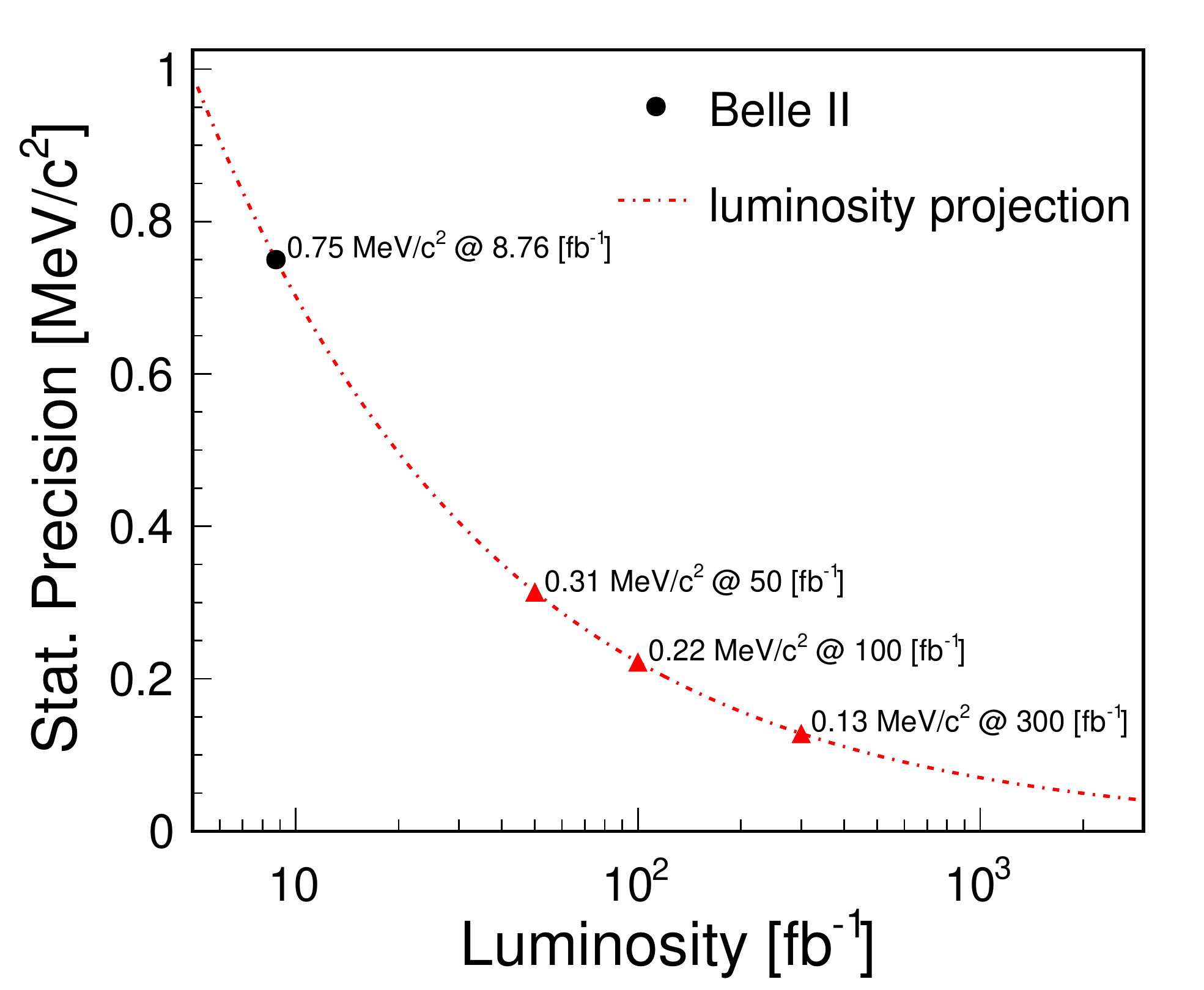}
    \caption{Projection of the statistical uncertainty as a function of luminosity for the $\tau$ mass measurement. The black dot represents the statistical uncertainty in this measurement and the red triangles mark the projected statistical uncertainties for 50, 100 and 300\invfb.}
    \label{fig:taumasbelle2projection}
\end{figure}

The leading source of systematic uncertainty is the momentum scale factor, which is expected to be reduced in the near future. 
With the present level of the systematic uncertainties, the Belle II $\tau$ mass measurement is expected to be statistically dominated until around 50\invfb of data.
After improvements in the momentum scale factor systematic uncertainty, a scenario with a total systematic uncertainty of 0.15\mevcc is foreseen, and about 300\invfb of data would be needed to become systematically dominated, as illustrated in Fig.~\ref{fig:taumasbelle2projection}. The systematic uncertainties can be reduced further by increasing the Monte Carlo statistics in the estimation of the bias, which is currently the second-largest systematic uncertainty. Thus, a better systematic precision is expected in the future. 

%% file: systematics.tex
The impact of various systematic sources of uncertainties on the $\tau$ mass measurement has been estimated.   Table~\ref{table:systsum} summarizes the systematic uncertainties; the sources are described in detail below. 
\begin{table}[h!]
\begin{center}
\begin{tabular}{lc }
\hline
Systematic uncertainty \; \; \; \; \; \; & \mevcc \\
\hline
\hline
Momentum shift due to the B-field map & \SystMomentumSF  \\
Estimator bias                        & \SystEstimator              \\ 
Choice of p.d.f.                      & \SystFitFunc                \\ 
Fit window                            & \SystWindow                 \\ 
Beam energy shifts                    & \SystBeamEnergy             \\
Mass dependence of bias               & \SystEstimatorMassDep       \\ 
Trigger efficiency                    & $\leq$ 0.01                  \\
Initial parameters                    & $\leq$ 0.01                 \\ 
Background processes                  & $\leq$ 0.01                      \\ 
Tracking efficiency                   & $\leq$ 0.01  \\
\hline
\end{tabular}
\end{center}
\caption{Summary of systematic uncertainties.}
\label{table:systsum}
\end{table}

\begin{itemize}
    \item  \textbf{Momentum shift due to the B-field map:} The leading source of uncertainty for this measurement comes from a momentum scale factor of $0.056_{-0.042}^{+0.051}\%$ that is introduced to compensate for the imperfections of the magnetic-field map used during the reprocessing of data. 
    The scale factor is measured according to an observed shift in the invariant mass of $D^{0}$ in data vs the PDG~\cite{Tanabashi:2018oca} value. The central value for the scale factor is used to correct the momenta of the tracks in data, and the average of the impact due to the up and down variations is used in Monte Carlo to estimate the associated systematic uncertainty of \SystMomentumSF~\mevcc.

   \item \textbf{Estimator bias:} The limited size of the samples used in determining the fit bias results in an uncertainty of 0.12~\mevcc. 
   
   \item \textbf{Dependence on the choice of p.d.f.:} Two alternate functions,
    
    \begin{align}
    \label{eq:pdf_f1}
    F_1(M,\vec{P}) &= (P_3 + P_4 M)\cdot\frac{M - P_1}
    {\sqrt{P_2+(M-P_1)^2}}+ P_5 M + 1, \\
    \label{eq:pdf_f2}
    F_2(M,\vec{P}) &= (P_3 + P_4 M)\cdot\frac{-1}{1+\exp{\big((M-P_1)/P_2}\big)} + P_5 M + 1,
   \end{align}
   
   are used for estimating the fit bias. The RMS
   of the corrected $\mtau$ values corresponding to these alternative methods versus the nominal one is calculated to be \SystFitFunc\mevcc.
   
    \item \textbf{Choice of the fit window:} The importance of the \Mmin window used for the fit procedure is tested by varying the lower and upper edges of the window separately and repeating the fit bias estimation for each case. The weighted average of the differences between the true mass and the corrected $\mtau$ values corresponding to each window is~\SystWindow~\mevcc.   
    
    \item \textbf{Beam energy shifts:} The calculation of the pseudomass variable relies on an accurate knowledge of the beam energy. 
    The energy of the beam is measured by using the beam-energy-constrained mass ($M_{\rm{bc}}$) of fully reconstructed neutral and charged B decays for various data taking periods with statistical uncertainties of up to 0.19~\mev. This uncertainty is then propagated to the $\tau$ mass measurement by taking advantage of additional Monte Carlo samples with beam energies shifted with respect to the nominal beam energy value. The measurement procedure is applied to each Monte Carlo sample to estimate the $\tau$ mass as a function of the beam energy shift. This yields a systematic uncertainty of \SystBeamEnergy~\mevcc. Additional systematic uncertainties in the measurement of the beam energy shift are not currently estimated. Once these uncertainties are available, they will be propagated to the $\tau$ mass measurement as well.

    \item \textbf{Dependence of the fit bias on the true mass:}
    Allowing for the fit bias to vary as a function of the true mass results in a difference of 0.02~\mevcc in the corrected mass with respect to the method described in Section~\ref{sec:measurement}. 

    \item \textbf{Trigger efficiency:} The trigger efficiency is measured to be an essentially constant value of $80\%$ in the pseudomass region used for the $m_{\tau}$ measurement. 
    The impact of this efficiency is seen to be negligible by repeating the fit on the signal Monte Carlo sample with a reweighting of the data by a linear parameterisation of the trigger efficiency as a function of $M_{min}$.
 
    \item \textbf{Model for background processes:} The distribution of background processes in the pseudomass window of interest is featureless. The fit procedure on the signal Monte Carlo sample is repeated by adding the background processes to the fit. The result of fit is insensitive to the presence of these background events.

    \item \textbf{
    Decay model of $a_1(1260)$:} To test the dependence of the fit result on the decay model of $\tau^+ \to  \pi^+\pi^-\pi^+\bar\nu_\tau$, the nominal fit is performed on the generated Monte Carlo sample with a phase space decay of the $a_1(1260)$.
    The shape of the $M_{min}$ distribution in the fit window is found to be independent of the $a_1(1260)$ decay model and therefore the size of this systematic is expected to be small.
    However, due to the large statistical uncertainties of the alternative decay scenario, at this point, a quantitative estimate of the size of this systematic uncertainty cannot be made. Within the uncertainty, the result of the fit on the alternative decay model is observed to be consistent with that of the nominal decay model \cite{decker1993tau}.
    
    \item \textbf{Tracking efficiency:} As a test, the track reconstruction efficiency is artificially reduced by $\sim 1 \%$ in Monte Carlo to match that in data. This reduction of the efficiency, however, has not impact on the result of the fit.

\end{itemize}

%% file: acknowledgements.tex
We thank the SuperKEKB group for the excellent operation of the
accelerator; the KEK cryogenics group for the efficient
operation of the solenoid; and the KEK computer group for
on-site computing support.
This work was supported by the following funding sources:
Science Committee of the Republic of Armenia Grant No. 18T-1C180;
Australian Research Council and research grant Nos.
DP180102629, 
DP170102389, 
DP170102204, 
DP150103061, 
FT130100303, 
and
FT130100018; 
Austrian Federal Ministry of Education, Science and Research, and
Austrian Science Fund No. P 31361-N36; 
Natural Sciences and Engineering Research Council of Canada, Compute Canada and CANARIE;
Chinese Academy of Sciences and research grant No. QYZDJ-SSW-SLH011,
National Natural Science Foundation of China and research grant Nos.
11521505,
11575017,
11675166,
11761141009,
11705209,
and
11975076,
LiaoNing Revitalization Talents Program under contract No. XLYC1807135,
Shanghai Municipal Science and Technology Committee under contract No. 19ZR1403000,
Shanghai Pujiang Program under Grant No. 18PJ1401000,
and the CAS Center for Excellence in Particle Physics (CCEPP);
the Ministry of Education, Youth and Sports of the Czech Republic under Contract No.~LTT17020 and 
Charles University grants SVV 260448 and GAUK 404316;
European Research Council, 7th Framework PIEF-GA-2013-622527, 
Horizon 2020 Marie Sklodowska-Curie grant agreement No. 700525 `NIOBE,' 
and
Horizon 2020 Marie Sklodowska-Curie RISE project JENNIFER2 grant agreement No. 822070 (European grants);
L'Institut National de Physique Nucl\'{e}aire et de Physique des Particules (IN2P3) du CNRS (France);
BMBF, DFG, HGF, MPG, AvH Foundation, and Deutsche Forschungsgemeinschaft (DFG) under Germany's Excellence Strategy -- EXC2121 ``Quantum Universe''' -- 390833306 (Germany);
Department of Atomic Energy and Department of Science and Technology (India);
Israel Science Foundation grant No. 2476/17
and
United States-Israel Binational Science Foundation grant No. 2016113;
Istituto Nazionale di Fisica Nucleare and the research grants BELLE2;
Japan Society for the Promotion of Science,  Grant-in-Aid for Scientific Research grant Nos.
16H03968, 
16H03993, 
16H06492,
16K05323, 
17H01133, 
17H05405, 
18K03621, 
18H03710, 
18H05226,
19H00682, 
26220706,
and
26400255,
the National Institute of Informatics, and Science Information NETwork 5 (SINET5), 
and
the Ministry of Education, Culture, Sports, Science, and Technology (MEXT) of Japan;  
National Research Foundation (NRF) of Korea Grant Nos.
2016R1\-D1A1B\-01010135,
2016R1\-D1A1B\-02012900,
2018R1\-A2B\-3003643,
2018R1\-A6A1A\-06024970,
2018R1\-D1A1B\-07047294,
2019K1\-A3A7A\-09033840,
and
2019R1\-I1A3A\-01058933,
Radiation Science Research Institute,
Foreign Large-size Research Facility Application Supporting project,
the Global Science Experimental Data Hub Center of the Korea Institute of Science and Technology Information
and
KREONET/GLORIAD;
Universiti Malaya RU grant, Akademi Sains Malaysia and Ministry of Education Malaysia;
Frontiers of Science Program contracts
FOINS-296,
CB-221329,
CB-236394,
CB-254409,
and
CB-180023, and SEP-CINVESTAV research grant 237 (Mexico);
the Polish Ministry of Science and Higher Education and the National Science Center;
the Ministry of Science and Higher Education of the Russian Federation,
Agreement 14.W03.31.0026;
University of Tabuk research grants
S-1440-0321, S-0256-1438, and S-0280-1439 (Saudi Arabia);
Slovenian Research Agency and research grant Nos.
J1-9124
and
P1-0135; 
Agencia Estatal de Investigacion, Spain grant Nos.
FPA2014-55613-P
and
FPA2017-84445-P,
and
CIDEGENT/2018/020 of Generalitat Valenciana;
Ministry of Science and Technology and research grant Nos.
MOST106-2112-M-002-005-MY3
and
MOST107-2119-M-002-035-MY3, 
and the Ministry of Education (Taiwan);
Thailand Center of Excellence in Physics;
TUBITAK ULAKBIM (Turkey);
Ministry of Education and Science of Ukraine;
the US National Science Foundation and research grant Nos.
PHY-1807007 
and
PHY-1913789, 
and the US Department of Energy and research grant Nos.
DE-AC06-76RLO1830, 
DE-SC0007983, 
DE-SC0009824, 
DE-SC0009973, 
DE-SC0010073, 
DE-SC0010118, 
DE-SC0010504, 
DE-SC0011784, 
DE-SC0012704; 
and
the National Foundation for Science and Technology Development (NAFOSTED) 
of Vietnam under contract No 103.99-2018.45.